\documentstyle[prb,aps,epsfig]{revtex}
\begin{document}

\title{ Interacting fermions in one dimension: The
  Tomonaga-Luttinger model }

\draft
\author { K. Sch\"onhammer}


\address{ Institut f\"ur Theoretische Physik der Universit\"at G\"ottingen
Bunsenstr. 9, D-37073 G\"ottingen, Germany}
\date{\today}
\maketitle
\begin{abstract}
 The theoretical description of interacting fermions in one
spatial dimension is simplified by the fact that the low energy
spectrum of noninteracting fermions is identical to the one for a
harmonic chain. This fermion-boson transmutation allows to describe
interacting fermions as a system of coupled oscillators.
Tomonaga's model of interacting
nonrelativistic fermions on a ring 
is presented and first discussed in low
order perturbation theory. 
After introducing the concept of two independent species of 
right and left moving fermions the exact
solution of the Tomonaga-Luttinger model is discussed in detail. The momentum
distribution and spectral functions are calculated using the method of
the bosonization of the field operator. The general Luttinger liquid
phenomenology is shortly discussed.
\end{abstract}

\section{Introduction}
The effects of the Coulomb interaction of electrons in metals can only
be described approximately \cite{AM,Ma}. With the progress in
producing artificial low dimensional structures the theoretical work
on {\em one-dimensional} interacting fermions has gained
importance \cite{V2}. Special features of the spectrum of low energy
excitations in one dimension allow {\em exact} solutions of
models of {\em interacting} fermions. The main idea can already
be understood by working with noninteracting fermions, which have the
same spectrum of excitations as a harmonic chain. This
``fermion-boson-transmutation'' (FBT) was discussed in a recent
publication on different levels of sophistication
\cite{SM}. In the following this paper is referred to as I. The basic
idea can be understood on the level of elementary quantum statistical
mechanics. Only in chapter V. of I the method of second quantization
was introduced to calculate the momentum distribution of
{\em noninteracting} fermions in the {\em canonical}
ensemble. It was pointed out that the technique presented can also be
used in the description of {\em interacting fermions}. From the
response we obtained to I it is clear that many readers would have
liked to see more explicitly the application to this problem. The
purpose of this paper is to fill this gap on a level which requires
only a basic knowledge of the method of second quantization but none
of relativistic quantum field theory.

The exact solution to the ``Tomonaga-Luttinger (TL-) model'' is
presented starting with Tomonaga's model of spinless interacting
nonrelativistic fermions on a ring \cite{T}. In section II the
momentum distribution in the interacting ground state is calculated in
leading order perturbation theory. This already demonstrates one of
the important ideas towards the exact solution and on the quantitative
level shows a sign of the infrared singularities later found in the
exact solution. Tomonaga's concept of ``right'' and ``left'' moving
electrons later extended by Luttinger \cite{L} is introduced and the
commutation relations of the Fourier components of the operator of the
particle density are derived. In order to avoid mathematical subtleties
we work with {\em two} cut-offs -- one for the number of
momentum states (``band cut-off'') and one for the inverse range of the
two-body interaction (``interaction cut-off''). This allows to
smoothly go from Tomanaga's original model to the model discussed by
Luttinger. In section III. the exact low energy eigenvalues of the
TL-model are obtained using the bosonization technique. As the most
important ``Luttinger-liquid'' (LL) feature the momentum distribution in
the ground state is calculated. Its derivative at the Fermi points
diverges in a power law fashion determined by the important constant
called ``anomalous dimension''.  One effort to experimentally verify
LL-properties is photoemission from quasi-one dimensional conductors
\cite{Da} . As the theoretical description of photoemission involves
 spectral functions they are discussed in section
IV. The model including spin is introduced in section V. After the
definition of ``charge'' and ``spin'' bosons the Hamiltonian of the
model can be written as a sum of two {\em commuting} terms of
the spinless type leading to the phenomenon of
``spin-charge-separation''. In section VI. the general LL-concept is
presented and it is shortly discussed how the LL parameters which
completely determine the low energy physics also for more complicated
$1d$ lattice models can be extracted from finite size numerical
data. In appendix A the proof of the Kronig-identity which provides
the bosonization of the kinetic energy is given. The bosonization of
the field operator is discussed in appendix B. A novel straightforward
approach is presented to determine the particle number changing
contribution.
We put $\hbar=1$ in this paper.

\section{Interacting fermions: The Tomonaga model}

The method of bosonization which was presented in I in the context of
noninteracting electrons in one dimension is the key concept to
understand ground state properties and the spectrum of excited states
with low excitation energy also for {\em interacting} fermions. In a
seminal paper Tomonaga \cite{T} studied nonrelativistic fermions in the
{\em high density} limit, in which the spatial range of the two-body
interaction is much larger than the interparticle distance. He studied
{\em spinless} fermions on a ring, i.e. using {\em periodic boundary
  conditions}. The Hamiltonian $H$ is a sum of the kinetic energy
$\hat T = \sum^{N}_{i=1}\hat p^2_i/2m$ and a two-body interaction
$\hat V$
\begin{eqnarray}
\hat V & = &
\frac{1}{2}\sum_{i \neq j}V(\hat x_i-\hat x_j)\\
& = &
\frac{1}{2}\int^{L/2}_{-L/2}dx\int^{L/2}_{-L/2}dx'
\psi^\dagger (x)\psi^\dagger (x')V(x-x')\psi (x')\psi(x),\nonumber
\end{eqnarray}

\noindent where in the second line we have switched to the language of second
quantization, with $\psi (x)$ the field operator which annihilates a fermion
at the position $x$. It can be expressed in terms of the
annihilation operators $c_n$ of momentum states in the standard way
\cite{Ma,B,warning}
\begin{equation}
\psi (x) = \frac{1}{\sqrt L}\sum_n e^{ik_nx}c_n,
\end{equation}

\noindent where the sum runs over momenta $k_n = 2\pi n/L, n \in \mathbf Z$
compatible with the periodic boundary condition. The $c_n (c^\dagger _m)$
obey canonical anticommutation relations
\begin{equation}
[c_m, c_n]_+ = 0 \; , [c_m, c_n^\dagger ]_+ = \delta_{mn}\hat 1.
\end{equation}

\noindent In order to be consistent with the periodic boundary
condition $\psi (x) = \psi (x+L)$ it is useful to expand the two body 
potential $V(x-y)$ in
a Fourier series
\begin{eqnarray}
V(x-y) &=&
\frac{1}{L}\sum_n e^{ik_n(x-y)}v(k_n),\\
v(k_n) &=&
\int^{L/2}_{-L/2}V(x)e^{-ik_nx}dx. \nonumber
\end{eqnarray}

\noindent Special choices of the
Fourier transformed potential $v(k)$ will be discussed later. In case
$V(0)$ exists, we can express $\hat V$ in terms of the operator of the
particle density
\begin{equation}
\hat \rho (x) = \sum^{N}_{i=1} \delta(x-\hat x_i) = \psi^\dagger (x)\psi (x),
\end{equation}

\noindent where the $\hat x_i$ are the position operators of the particles. 
Using $[\psi(x), \psi^\dagger (y)]_+ = \delta(x-y)$ for $x,y \in [-L/2,L/2]$
one obtains
\begin{equation}
V = \frac{1}{2}\int^{L/2}_{-L/2}dx\int^{L/2}_{-L/2}dx' V(x-x')\hat
\rho(x)\hat\rho(x') - \frac{1}{2}V(0){\cal N},
\end{equation}

\noindent where we have introduced the particle number operator ${\cal
  N}\equiv \int^{L/2}_{-L/2}\rho (x)dx$. As in Tomonaga's paper we
  present the main results for the spinless model and discuss the
  modifications due to the electron spin later. It is
useful to Fourier decompose the density operators
\begin{eqnarray}
\hat\rho(x) &=&
\frac{1}{L}\sum_n \hat\rho_n e^{ik_nx},\\
\hat\rho_n &=&
\int^{L/2}_{-L/2} \hat\rho (x)e^{-ik_nx}dx.  \nonumber
\end{eqnarray}

If we use the second quantized version $\hat \rho (x) = \psi^\dagger (x)\psi
(x)$ the $\hat \rho_n$ can be expressed as follows
\begin{equation}
\hat \rho_n = \sum_{n'} c^\dagger _{n'}c_{n'+n}.
\end{equation}

In terms of the $\hat \rho_n$ the interaction term is given by
\begin{equation}
\hat V = \frac{1}{L}\sum_{n>0} v(k_n)\hat  \rho_n \hat \rho_{-n}  +
\frac{1}{2L}{\cal N}^2 v(0) - \frac{1}{2} V(0){\cal N}.
\end{equation}

\noindent We have used $v(k_n) = v(-k_n)$ as $V(x)$ is real and $[\hat \rho_n,
\hat \rho_m] = 0$, which follows most easily from 
the fact that only the commuting
position operators $\hat x_i$ enter in the definition of $\hat \rho
(x)$ in Eq. (5).

Now following Tomonaga we assume that the spatial range of the
two-body interaction is much larger than the mean particle separation
which is inversely proportional to the Fermi momentum $k_F$. Two
examples for $v(k)$ with a finite value of $v(0)$
which will be used later in the paper
are shown in Fig. 1.
\begin{figure} [hbt]
\hspace{4.0cm}
\vspace{1.0cm}
\epsfig{file=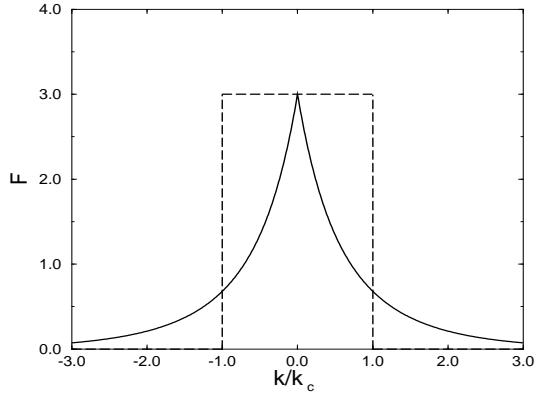,width=6cm,height=8cm,angle=-90}
\caption{ Two forms of the dimensionless 
Fourier transform $F(k)\equiv v(k)/(\pi v_F)$ of the two body
potential. }
\end{figure}
The ground state of the noninteracting system with $N$ fermions ($N$ odd)
is the Slater determinant $|F>$ (''Fermi sea'') in which the
momentum states from $-k_F = -2\pi n_F/L$ to $k_F$ are occupied, where
$n_F = (N-1)/2$ and $k_F$ is the Fermi momentum. 
 A first insight into the nature of the {\em
  interacting} ground state can be obtained using first order
perturbation theory in $\hat V$ \cite{B}
\begin{equation}
|\phi_0> = |F> + \frac{\hat Q}{E_0^{(0)} - \hat T} \hat V|F> + ...,
\end{equation}
where $\hat Q=1-|F><F|$ and $E^{(0)}_0$ is the
kinetic energy of the Fermi sea. Due to the projector $\hat Q$ only
the first term on the right hand side (rhs) of Eq. (9) 
contributes to the first
order correction in the above equation. As we assume $v(k_n)$ to be
zero for $k_n$ (much) larger than a cut-off $k_c \ll k_F$ only
states $\hat \rho_n \hat \rho_{-n}|F>$ with $n$ of the order $n_c =
k_cL/2\pi$ or smaller occur in $\hat V |F>$. The second quantized form
of $\hat \rho_n$ Eq. (8) shows that $\hat \rho_n \hat \rho_{-n}|F>$ is a
{\em two particle-two hole state}. For $n_c >  n>0$ the
density operator $\hat\rho_{-n}$ creates a particle-hole pair 
around the right Fermi point $k =
k_F$, while $\hat \rho_n$ creates a particle-hole pair with opposite
momentum around the left Fermi point $k = -k_F$
\begin{equation}
\hat Q \hat V |F> = \frac{1}{L} 
\sum_{n>0}v(k_n)
\sum^{-n_F+1}_{n''=-n_F-n}\;
\sum^{n_F+n}_{n'=n_F+1} 
c^\dagger _{n''}\;c_{n''+n}\;c^\dagger _{n'}\;c_{n'-n}|F>
\end{equation}
The kinetic energy of the two-particle-two hole state
$c^\dagger _{n''}\;c_{n''+n}\; c^\dagger _{n'}\;c_{n'-n}|F>$ 
is given by $E_0^{(0)}+
\epsilon_{n''}-\epsilon_{n''+n}+\epsilon_{n'}-\epsilon_{n'-n}$ where
$\epsilon_n = (2\pi n/L)^2/2m_e$. As the particles and the holes
are in the {\em vicinity} of the Fermi points one can linearize the
energy dispersion around the two Fermi points in order to calculate
$\epsilon_{n'} - \epsilon_{n'-n}$ and $\epsilon_{n''} -
\epsilon_{n''+n}$. With the Fermi energy $\epsilon_F =
k^2_F/2m_e$ and the Fermi velocity $v_F = k_F/m_e$ one uses
instead of the {\em one} parabolic dispersion {\em two} linear branches
\begin{eqnarray}
\epsilon_{n,+} & \equiv &
\epsilon_F + v_F (n-n_F)2 \pi/L \quad n > 0\\ \nonumber
\epsilon_{n,-}& \equiv &
\epsilon_F - v_F (n+n_F)2\pi/L \quad n < 0 
\end{eqnarray}
As will be discussed later the use of the linear dispersions, also
shown in Fig. 2, leads to an error which vanishes in the 
high density limit
$k_F/k_c \to \infty$.
 It simplifies the expression for the kinetic
energy of the two particle-two hole state to $E_0^{(0)} + 2 k_n v_F$ 
and leads to
\begin{equation}
|\phi_0> = |F> - \frac{1}{L}\sum_{n>0}\frac{v(k_n)}{2k_nv_F}
\;\sum^{-n_F+1}_{n''=-n_F-n}\;
  \sum^{n_F+n}_{n'=n_F+1}c^\dagger _{n''}\;c_{n''+n}\;
c^\dagger _{n'}\;c_{n'-n}|F> +
  ...
\end{equation}
Higher order perturbation theory yields additional terms with multiple
particle-hole pairs.\\

An important way to characterize the interacting ground state is to
calculate its momentum distribution

\noindent  $<\phi_0|\hat n_m|\phi_0> \equiv
<\phi_0|c^\dagger _mc_m|\phi_0>$, i.e. to study how the Fermi step
distribution of the noninteracting electrons is modified by the
interaction. It follows from Eq. (13) that the leading correction to
the Fermi function is quadratic in the two-body potential. For $m>n_F$
one obtains e.g. 
\begin{equation}
<\hat n_m>^{(2)} = \frac{1}{L^2}\sum_{n \ge m-n_F}\;\;\sum^{-n_F+1}_{n'' =
  -n_F-n}
\left(\frac{v(k_n)}{2k_n v_F}\right)^2,
\end{equation}
where we have used that the two particle-two hole states are
eigenstates of $\hat n_m$. The sum over $n''$ can be trivially
performed and yields a factor $n$. The sum over $n$ is especially
simple for the step potential $v(k) = v(0) \Theta \left(
  k^2_c-k^2\right)$ where $\Theta (\cdot)$ is the unit step function. If
the sum over $1/n$ is approximated by an integral one finally obtains
\begin{equation}
k_m > k_F : \quad <\hat n_m>^{(2)} = - \left( \frac{v(0)}{4\pi
v_F}\right)^2 \ln \left(\frac{k_m -k_F}{k_c}\right) \cdot \Theta
\left[
k_c-\left(k_m-k_F\right)\right] . 
\end{equation}
The logarithmic singularity for $(k_m-k_F)/k_c \to 0$ is an artefact
of the perturbation expansion, as the occupancies have to be smaller
than one. The exact solution for the momentum distribution $<\hat
n_m>$ will be derived in detail later. In the thermodynamic limit
$L \to \infty, N/L = const$ one obtains a continuous function $n(k)$
which in the neighbourhood of $k_F$ is given by 
\begin{equation}
n(k) = \frac{1}{2} + c \left| \frac{k-k_F}{k_c}\right|^\alpha \mbox{
sign}\; \left(k_F - k \right).
\end{equation}
The constant $\alpha$ is called the {\em anomalous dimension}. For
weak two-body potentials $\alpha$ is small and one can use $x^\alpha =
\exp(\alpha \ln x) = 1 + \alpha \ln x + O(\alpha^2)$ for fixed $x$. By
comparison with the perturbational result Eq. (14) one finds $c = 1/2$
and 
\begin{equation}
\alpha^{(2)} = \frac{1}{2} \left(\frac{v(0)}{2\pi v_F}\right)^2
\end{equation}
as the lowest order perturbational result for the anomalous
dimension. A quantitative comparison of $<\hat n_m>^{(2
)}$ with the
exact result will be given later.\\

Perturbation theory shows that there are three types of electrons:
Those with momenta $k \approx k_F$, which move with velocities
$\approx v_F$ which will be called ''right movers'', those with
momenta $k \approx -k_F$ and velocities $\approx -v_F$ (''left
movers'') and the ''inert'' electrons deep in the Fermi sea, which
play no role in the ''low energy physics''. This basic insight of
Tomonaga was used by Luttinger \cite{L} to further modify the model. In
addition to the linearization Eq. (12) he extended the range of the
possible $k$-values for {\em both} the right and left movers to $2\pi
n/L$ with $n$ positive and negative integers. This
further simplifies the dispersion of the electrons, but introduces
infinitely many negative energy electron states, which have to be filled in
the ground state. As this infinite ''Dirac sea'' leads to various
mathematical subtleties we prefer to work with a {\em finite band
cut-off}.
We introduce the momentum  $k_B=2\pi m_0/L\geq 0$ and allow states
with momenta $\geq -k_B$ for right movers and $\leq k_B$ for left 
movers. In Fig.2 we have chosen $k_B=1.5k_F$ and show the states added
as  the dot-dashed lines.

\begin{figure} [hbt]
\hspace{4.0cm}
\vspace{1.0cm}
\epsfig{file=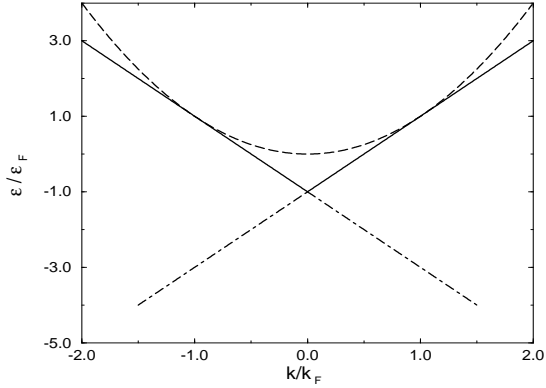,width=6cm,height=8cm,angle=-90}
\caption{Energy dispersion as a function of momentum. The dashed curve
shows the usual ``nonrelativistic'' dispersion used in the original 
Tomonaga model and the full curve the linearized version Eq. (12).
The dot-dashed parts are the additional states for $k_B=1.5k_F$. The 
model discussed by Luttinger corresponds to $k_B \to \infty$.}
\end{figure}
Later we also consider the limit $k_B \to \infty$. 
The kinetic energy of the model
solved exactly in the next section reads

\begin{equation}
\hat T_{TL} = \sum_{n\in I_{+}} v_F k_n c^\dagger _{n,+}c_{n,+} + \sum_{n \in
  I_{-}}\left(-v_F k_n\right) c^\dagger _{n,-} c_{n,-}\;,
\end{equation}
where $I_+ = \{-m_0, -m_0+1, ...\infty\}$ and $I_- = \{-\infty,
... m_0-1, m_0\}$. The creation and annihilation operators for the right
$(\alpha = +)$ and left $(\alpha = -)$ movers obey the fermion
anticommutation relations $(m \in I_\alpha, n \in I_\beta)$
\begin{equation}
\left[c_{m,\alpha}, c_{n,\beta}\right]_+ = 0\; \quad
\left[c_{n,\alpha}, c^\dagger _{m,\beta}\right]_+ = 
\delta_{n,m}\delta_{\alpha,\beta}\hat 1.
\end{equation}
In the interaction term $\hat V$ Eq. (9) we replace $\hat \rho_n 
\hat \rho_{-n}$
by $\left(\hat \rho_{n,+} +\hat \rho_{n,-}\right)\left( \hat \rho_{-n,+} +
 \hat \rho_{-n,-}\right)$ where
\begin{equation}
\hat \rho_{n,\alpha} \equiv \sum_{n'} w^\alpha_{n',n'+n} \; c^\dagger _{n',\alpha}
\; c_{n'+n,\alpha}, 
\end{equation}
with
$$
w^\alpha_{n',n'+n} = \left\{ \begin{array}{c @{\hspace{2ex}}l}
1 & {\rm for}\; n', n'+n \in I_\alpha \\
0 & {\rm else}\end{array} \right. \nonumber
$$
are the ''density operators'' of the right and left movers. 
Because $w^\alpha_{n,m}$ is symmetric in $n$ and $m$ 
the relation $\hat \rho^\dagger _{m,\alpha} =
\hat \rho_{-m,\alpha}$ holds.

It is not more difficult to work with the following generalization
\begin{eqnarray}
\hat V_{TL}
& = & \frac{1}{L} \sum_{n>0}
\left\{g_4(k_n)
\left(\hat \rho_{n,+}\;\hat \rho_{-n,+} +\hat \rho_{n,-}\;
\hat \rho_{-n,-}\right)
+ g_2 (k_n)\left(\hat \rho_{n,+}\;\hat \rho_{-n,-} +\hat \rho_{n,-}\;
 \hat \rho_{-n,+}\right)\right\}\nonumber \\
& &
+ \frac{1}{2L}g_4(0)
\left( {\cal N}^2_+ + {\cal N}^2_- \right) +
\frac{1}{2L}g_2(0)\cdot 2{\cal N}_+ {\cal N}_-
\end{eqnarray}
where the ${\cal N}_\alpha$ are the particle number operators
\begin{equation}
{\cal N}_\alpha \equiv \sum_{m \in I_\alpha} c^\dagger
 _{m,\alpha}\; 
c_{m,\alpha}
\end{equation}
and we have dropped the terms proportional to the particle number
operators as they can be later included in the chemical potential . 
The second line in Eq. (21) can also be expressed in terms of
the total particle number operator ${\cal N}\equiv {\cal N}_+ +
{\cal N}_-$ and the ''current operator'' ${\hat J} \equiv {\cal N}_+ -
{\cal N}_-$ \cite{H}.
 For the interaction functions in Eq. (21) we have used the
standard notation \cite{V2}. 
 If we put $g_4(k)\equiv g_2(k)= v(k)$ the
Tomonaga-Luttinger Hamiltonian $H_{TL} \equiv \hat T_{TL} + \hat
V_{TL}$ yields in perturbation theory in $\hat V_{TL}$ the {\em same
  result}
 for $<\hat n_{m,+}>$ as in Eq. (15). In $\hat Q\; \hat V_{TL}|F_{TL}>$
only the terms proportional to $\hat \rho_{n,-}\;
\hat \rho_{-n,+}|F_{TL}>$ contribute
for $n>0$. Concerning the low energy physics the original model with
the quadratic dispersion is equivalent to the {\em Tomonaga-Luttinger
  (TL) model} described by the Hamiltonian $H_{TL}$ in the high density
limit ${k_F/k_c}\to \infty$.

There is an essential difference in the
commutation relations of the $\hat \rho_{m,\alpha}$ and the $\hat
\rho_m$. 
While
as discussed previously $[\hat \rho_m,\hat \rho_n] = 0$, which can be seen most
easily in the first quantized version, the $\hat \rho_{m,\alpha}$ do {\em
  not} all commute. From the anticommutation relations for the
fermions on different branches it follows that $[\hat
\rho_{m,+},
\hat \rho_{n,-}] = 0$. The commutators $[\hat \rho_{m,\alpha}, 
\hat \rho_{n,\alpha}]$
on the other hand are nontrivial
\begin{eqnarray}
[\hat \rho_{m,\alpha},\hat \rho_{n,\alpha}]
& = &
\sum_{m',n'}
[c^\dagger _{m',\alpha} \; c_{m'+ m,\alpha},\; c^\dagger _{n'\alpha} \; 
c_{n'+ n,\alpha}]
w^\alpha_{m',m'+m} w^\alpha_{n',n'+n}\\ \nonumber
& = &
\sum_{m'n'}
\left(\delta_{n',m'+m}\; c^\dagger _{m',\alpha}\; c_{n'+n,\alpha} 
-\delta_{m',n'+n}\;  c^\dagger _{n'\alpha}\; c_{m'+m,\alpha}\right)
w^\alpha_{m',m'+m}\; w^\alpha_{n',n'+n}\\ \nonumber
& = &
\sum_{m'}
\left( w^\alpha_{m',m'+m} - w^{\alpha}_{m',m'+n} \right)
w^\alpha_{m', m'+m+n}\; c^\dagger _{m',\alpha} \; c_{m'+m+n}.
\end{eqnarray}
In going to the last equation the summation index was changed from $n'$
to $m'$ in the second term. It is easy to see that the rhs of Eq. (23)
{\em vanishes} if $m$ and $n$ have the {\em same} sign. Take, for
example, right movers and $m$ and $n$ positive. Then due to the factor
$w^+_{m',m'+m+n}$ the rhs of Eq. (21) can only be nonzero for $m' \ge
- m_0$ and $m'+ m + n \ge - m_0$. But then also $m'+m \ge -m_0$ {\em and}
$m' + n \ge - m_0$ and the commutator vanishes because of the factor 
$w^+_{m',m'+m} - w^+_{m', m'+n}= 0$. Next we consider the case $n = -m$
shortly discussed in section V. in I. For $ m > 0$ one obtains for the
right movers
\begin{equation}
[\hat \rho_{m,+},\hat \rho_{-m,+}] = \sum^{-m_0+m-1}_{m' = -m_0}
 c^\dagger _{m',+} \;
c_{m',+} \equiv  {\cal N}_+(m).
\end{equation}
We see that the commutator is given by the particle number of the
lowest $m$ right moving one-particle states. As explained earlier an
interaction $v(k)$ which drops off rapidly for $|k|>k_c$ does not
produce holes deep in the Fermi sea in the interacting ground state
and eigenstates with low excitation energy. In this subspace of the
total Hilbert space it is an excellent approximation which becomes
 asymptotically exact for $m_0 \to \infty$ to
replace ${\cal N}_+(m)$ by $m \hat 1$ on the rhs of Eq. (24). In the
interaction term $V_{TL}$ Eq. (20) only $\hat \rho_{n,\alpha}$ 
for $|n|\ll n_F$ contribute. Therefore it is sufficient to discuss the
commutation relations Eq. (23) for $m$ and $n$ with different sign for
$|m|, |n| \ll n_F$. For $m > -n > 0$ one obtains for right movers
\begin{equation}
[\hat \rho_{m,+},\hat \rho_{-n,+}] = \sum^{-m_0-n-1}_{m' = -m_0}
c^\dagger _{m',+} \; c_{m'+m-n,+} = -\sum^{-m_0-n-1}_{m'= -m_0}
c_{m'+m-n,+}\;
c^\dagger _{m',+}.
\end{equation}
The operator $c^\dagger 
_{m',+}$ tries to create an electron deep in the
Fermi-Dirac sea, which is not possible in the space of low lying
excited states, as all states at the bottom of the band are
occupied. Therefore the commutator vanishes in this
subspace. Analogous arguments apply to the left movers and we finally
obtain 
\begin{equation}
[\hat \rho_{m,\alpha},\hat \rho_{n,\beta}] = \alpha m \,\hat 1 \;\delta_{\alpha
  \beta}\;\delta_{m,-n}.
\end{equation}
The importance of this relation was first realized by Tomonaga \cite{T}
who worked with $k_B=0$. Luttinger \cite{L} made an error with the
commutation relation for $k_B=\infty$ 
later corrected by Mattis and Lieb \cite{ML}.

\section {Exact solution of the TL-model using bosonization}

The commutation relations Eq. (26) for the operators $\hat \rho_{m,\alpha}$
apart from a proper normalization factor, look like boson commutation
relations. If we define the operators
\begin{equation}
b_n \equiv \frac{1}{\sqrt{|n|}} 
\left\{ \begin{array}{c @{\hspace{3ex}}l}
\hat \rho_{n,+} & {\rm for}\; n > 0 \\     
\hat \rho_{n,-} & {\rm for}\; n < 0 \end{array} \right. 
\end{equation}
and the corresponding adjoint operators $b^\dagger _n$ using
$\hat \rho^\dagger_{n,\alpha} =\hat \rho_{-n,\alpha}$ Eq. (26) 
implies the usual
bosonic commutation relations
\begin{equation}
[b_n, b_m] = 0, \quad [b_n, b^\dagger _m] = \delta_{mn}\hat 1.
\end{equation}
The kinetic energy $\hat T_{TL}$ in Eq. (18) consists of two commuting
terms, which both are of the form discussed for fixed boundary
condition in I. Therefore they both can be expressed in terms of the
boson operators with the help of the Kronig identity \cite{Kro}
which was stated
in I without giving an explicit proof. This is therefore presented in
appendix A. It yields
\begin{equation}
\hat T_{TL} = \sum_{n\neq 0} v_F \left(\frac{2\pi}{L}\right)|n| b^\dagger _n b_n +
v_F \left( \frac{\pi}{L}\right)\left( {\cal N}^2_+ + {\cal
    N}^2_-\right) + {\rm const.} \; {\cal N}
\end{equation}
As we dropped terms proportional to ${\cal N}$ in $\hat V_{TL}$ which
only 
lead to a renormalization of the chemical
  potential we also drop the term proportional to ${\cal N}$ in
  the kinetic energy in the following. Neglecting also a constant term
  in the total Hamiltonian it reads \cite{H}
\begin{eqnarray}
H_{TL}
& = &
\sum_{n>0} \left\{k_n \left(v_F +\frac{g_4(k_n)}{2\pi}\right)
\left( b^\dagger _n b_n + b^\dagger _{-n}b_{-n}\right) +
\frac{k_n g_2 (k_n)}{2\pi}\left(b^\dagger _n b^\dagger _{-n} +
 b_{-n}b_n \right)
\right\}\nonumber\\
& &
+ \frac{\pi}{2L}\left[ v_N {\cal N}^2 + v_J {\cal J}^2 \right]
\equiv H_B + H_{{\cal N}, \hat {\cal J}}
\end{eqnarray}
with the velocities
\begin{eqnarray}
v_N & = &
v_F + \frac{g_4 (0) + g_2(0)}{2 \pi}\\ \nonumber
v_J & = & 
v_F + \frac{g_4 (0) - g_2(0)}{2 \pi}.
\end{eqnarray}
Here $v_N$ determines the energy change for adding particles while
$v_J$ enters the energy change when the difference in the number
of right and left movers is changed.
 For our original model with $g_4(q) = g_2(q) =
v(q)$ the velocity $v_J$ equals the Fermi velocity $v_F$ of the
noninteracting system. As the particle number operators $ {\cal
  N}_\alpha$ commute with the boson operators $b_m(b^\dagger _m)$ the two terms
$H_B$ and $H_{{\cal N},\hat {\cal J}}$ in the Hamiltonian commute
and can be treated separately.

The boson part $H_B$ of the $TL$-Hamiltonian is bilinear in the boson
creation and annihilation operators. As shown below it can be brought
into the form
\begin{equation}
H_B = \sum_{n \neq 0} \omega_n \alpha^\dagger _n \alpha_n + {\rm const.}
 \end{equation}

\noindent by solving an eigenvalue problem, and the $\alpha^\dagger
_n(\alpha_n)$ 
are also
boson creation (annihilation) operators. As $H_B$ in Eq. (30) is a sum of
commuting terms $H_{B,n}$ only the coupling of the four operators
$b_n, b_{-n}, b^\dagger _n, b^\dagger _{-n}$ has to be considered at a
time. Momentum conservation further reduces the $4 \times 4$
eigenvalue problem to a $ 2 \times 2$ problem. The eigenvalue problem
is obtained by looking at the Heisenberg equations of motion for
$b_n(t) = e^{iH_Bt} b_n e^{-iH_Bt}$. This involves the commutators
\begin{eqnarray}
\left( \begin{array}{ccc}
\left [b_n,H_B\right ]\\
\lbrack b^\dagger _{-n},H_B \rbrack \\
\end{array} \right )
=\left( \begin{array}{ccc}
k_n(v_F+g_4(k_n)/(2\pi )) & k_ng_2(k_n)/(2\pi )\\
-k_ng_2(k_n)/(2\pi ) & -k_n(v_F+g_4(k_n)/(2\pi))\\
\end{array} \right )
\left( \begin{array}{ccc}
b_n\\
b^\dagger _{-n}\\
\end{array} \right )
\end{eqnarray}
The matrix ${\bf M}^{(n)}$ on the rhs of Eq. (33) has the 
eigenvalues
\begin{equation}
\omega^{\pm}_n = \pm v_F k_n \sqrt{\left( 1 + \frac{g_4 (k_n)}{2 \pi
      v_F}\right)^2 -
\left( \frac{g_2(k_n)}{2\pi v_F}\right)^2} 
\equiv \pm \omega_n
\end{equation}
As ${\bf M}^{(n)}$ is not symmetric one has to distinguish
left and right eigenvectors. We denote the left eigenvector to
$\omega^+_n$ as $(c_n,-s_n)$ where $c_n$ and $s_n$ can be chosen as real
quantities as ${\bf M}^{(n)}$ itself is real. The formal
solution of the Heisenberg equations of motion $i\dot{b}_n = [b_{n},H]$
etc. reads
\begin{equation}
\left(\begin{array}{l}
b_n(t)\\
b^\dagger _{-n}(t)\end{array}\right) 
= e^{-i {\bf M}^{(n)}t}
\left(\begin{array}{l}
b_n\\
b^\dagger _{-n}\end{array}\right) .
\end{equation}
If we multiply this equation form the left with $(c_n, -s_n)$ we
can replace the matrix by its eigenvalue
  $(c_n,-s_n)e^{-i {\bf M}^{(n)}t} = (c_n, -s_n)
e^{-i\omega_nt}$ 
which shows that the operator 
\begin{equation}
\alpha_n \equiv c_n b_n - s_n b^\dagger _{-n}
\end{equation}
has the simple time evolution $\alpha_n (t) = e^{-i\omega_n
  t}\alpha_n$. With a proper normalization of the left eigenvector the
  $\alpha_n$ are therefore the boson operators which ''diagonalize''
  $H_B$ to the form Eq. (32). The $\alpha_n$ should obey the boson
  commutation relation $[\alpha_n, \alpha^\dagger_n] =\hat 1$. As Eq. (36)
  yields $[\alpha_n,\alpha^\dagger _n] = c^2_n - s^2_n$ the left eigenvector
  $(c_n,-s_n)$ has to be normalized by $c^2_n - s^2_n = 1$. Elementary
  calculation of the eigenvector then yields
\begin{equation}
s^2_n = \frac{1}{2}\left[
\frac{1 + g_4(k_n)/(2\pi v_F)}
{\sqrt{(1 + g_4(k_n)/(2\pi v_F))^2 - (g_2(k_n)/(2\pi v_F))^2}} -
1\right]
= s^2_{-n}
\end{equation}
In the noninteracting limit $g_\nu \to 0$ the weights $s_n$ vanish and
the ''Bogoljubov transformation'' in Eq. (36) reduces to $\alpha_n =
b_n$. The dispersion relation $\omega_n^{(0)} = v_F|k_n|$ of the
bosons in the noninteracting case (see I) is in the interacting case
replaced by $\omega_n = v_c(k_n)|k_n|$ where the $k$-dependent ''sound
velocity'' usually called the ''charge velocity'' follows from
Eq. (34) as
\begin{equation}
v_c(k_n)= v_F\sqrt{\left(
1 + \frac{g_4(k_n)+ g_2(k_n)}{2\pi v_F}\right)
\left(1+ \frac{g_4(k_n) - g_2(k_n)}{2 \pi v_F}\right)}.
\end{equation}
For our original model $g_4(k)\equiv g_2(k)\equiv v(k)$ it is given by
$v_c(k_n) = v_F(1 + v(k_n)/(\pi v_F))^{1/2}$. The corresponding
dispersion is shown in Fig. 3 for the two potentials $v$ presented in
Fig. 1.
\begin{figure} [hbt]
\hspace{4.0cm}
\vspace{1.0cm}
\epsfig{file=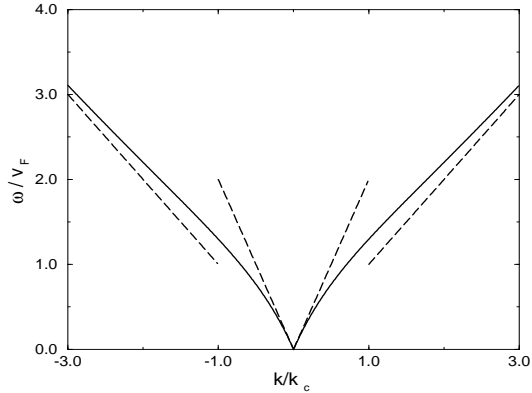,width=6cm,height=8cm,angle=-90}
\caption{  Dispersion of the bosonic modes in Eq. (34) for the two
potentials
shown in Fig. 1. For the potential which is constant up to $k_c$ 
the dispersion is discontinous while for the smooth potential 
the approach to the noninteracting limit for large $k$ is gradual.  }
\end{figure}
 For $|k_n|/k_c\ll 1$ the boson dispersion is approximately
linear $\omega_n \approx v_c|k_n|$ with $v_c \equiv v_c(0)$. From
Eqs. (31) and (38) we obtain the relation
\begin{equation}
v_c = \sqrt{v_N v_J}
\end{equation}
first pointed out by Haldane \cite{H}. For large momenta $|k_n|\gg k_c$
the dispersion of the noninteracting case is approached. In the figure
we have used a {\em purely} repulsive interaction $(v(k)\ge 0)$ which
leads to $v_c(k)\ge v_F$. For attractive interaction there is an
instability if $-v(q)$ gets larger than $\pi v_F$ \cite{ML}, leading to
phase separation. In the following we concentrate on repulsive
interactions.

The interacting ground state $|\{0\}_\alpha, \tilde N_+,\tilde N_->$
with the integers $\tilde N_\alpha \equiv N_\alpha - m_0 - 1$ is determined by 
$\alpha_n|\{0\}_\alpha,\tilde N_+,\tilde N_- > = 0 $. It can be
constructed by a unitary transformation out of the noninteracting ground state
\begin{equation}
|\{0\}_b, \tilde N_+, \tilde N_- > = \left(
 \prod^{\tilde N_-}_{n=-m_0} c^\dagger _{-n,-}\right) \left(
 \prod^{\tilde N_+}_{l= -m_0} c^\dagger _{l,+}\right)|{\rm Vac}\; >.
\end{equation}
Here we have allowed the number of right and left movers to be
different. Our original model corresponds to $\tilde N_+ = \tilde N_-
= n_F$. In the following we calculate expectation values in the
interacting ground state by algebraic methods using
$\alpha_n |\{0\}_\alpha, \tilde N_+, \tilde N_- > = 0$. Therefore the
explicit form of this state \cite{ML}
is not needed. The excited states for
given particle numbers $N_+, N_-$ are constructed as for harmonic
oscillators
\cite{B}
\begin{equation}
|\{m_l\}_\alpha, \tilde N_+, \tilde N_-> = \prod_{l \neq
  0}\frac{(\alpha_l^\dagger )^{m_{l}}}{\sqrt{m_l!}}|\{0\}_\alpha, \tilde N_+,
  \tilde N_- >.
\end{equation}
Their energy is given by $E_0(\tilde N_+, \tilde N_-) + \sum_{l\neq 0}m_l
\omega_l$. 

The nature of the interacting ground state shows up clearly in the
momentum distribution $<\hat n_{k_{n,\alpha}}>$. In the following we
calculate $<\hat n_{k_{n,+}}>$ and show how the Fermi step of the
noninteracting limit is modified near the right Fermi point $k_F$. As
in I it can be obtained by first calculating $<\tilde \psi_+^\dagger (u)
\tilde \psi_+ (v)>$, where $ \tilde \psi_+ (v) $ is the auxiliary
field operator defined in (B.6)
and then using the inversion formula (B.24) to
calculate the momentum distribution. 

With the use of the bosonization formula (B.25) the calculation is
quite simple. As we want to use $\alpha_n |\{0\}_\alpha, \tilde N_+,
\tilde N_- > = 0$
we express in the operator function $\phi_+$ defined in (B.9) the $b_n$
in terms of the $\alpha_n, \alpha^\dagger _{-n}$. From $\alpha_n = c_n b_n -
s_n b^\dagger_{-n}$ in Eq.
 (36)
one obtains the inversion 
\begin{equation}
b_n = c_n\alpha_n + s_n \alpha^\dagger _{-n}.
\end{equation}
This yields

\begin{equation}
i(\phi_+(u) - \phi_+(v)) = \sum^\infty_{n=1} \frac{e^{inu} -
  e^{inv}}{\sqrt{n}}(c_n \alpha_n + s_n \alpha^\dagger _{-n}).
\end{equation}
As the $\alpha_n$ and $\alpha^+_{-n}$ commute, $\exp [-i(\phi_+(u)- \phi_+
(v))]$ can be written as a product of two exponentials with the
annihilation part to the right. If applied to the interacting ground
state this factor yields unity, i.e.
\begin{equation}
e^{-i(\phi_+(u)-\phi_+(v))}|\{0\}_\alpha, \tilde N_+, \tilde N_-> =
e^{-\sum^{\infty}_{n=1}\left(
    \frac{e^{inu}-e^{inv}}{\sqrt{n}}\right)s_n
  \alpha^\dagger _{-n}}|\{0\}_\alpha, \tilde N_+, \tilde N_->.
\end{equation}
The expectation value of $\tilde\psi_+^\dagger(u)\tilde\psi_+(v)$ as seen from
Eq. (B.25) involves the overlap of the above state with
itself. Using the Baker Hausdorff formula $e^Ae^B=e^Be^Ae^{[A,B]}$ we
obtain
\begin{equation}
<\tilde \psi^\dagger_+(u) \tilde \psi_+(v)> = \frac{e^{-i\tilde
    N_{+}(u-v)}}{1-e^{i(u-v+i0)}} \exp\left\{-\sum^{\infty}_{n=1}
    \frac{2s_{n}^{2}}{n}(1-\cos [(u-v)n])\right\}.
\end{equation}
This is exactly of the form as the {\em finite temperature} canonical
expectation value for {\em noninteracting} fermions (Eq.(50) in I)
but with the Bose function $b(n\Delta\beta)$ replaced by $s^2_n$ which
depends on the interaction as shown in Eq.(37). For arbitrary
potentials therefore $<\hat n_{m,+}>$ can be calculated by the
recursive method outlined in the appendix of I. The sum in the
exponent on the rhs of Eq. (45) is especially simple, if $s^2_n$
decays exponentially
\begin{equation}
s^2_n = s^2(0) e^{-2|k_n|/k_c}.
\end{equation}
The corresponding potential $v(k_n)$ can be obtained using Eq. (37) by
solving a quadratic equation. The factor $2$ in the exponent on the
rhs of Eq. (46) was introduced because $v(k_n)$ is proportional to
$s_n$ for large momenta. 
 It is shown as the full line in Fig. 1
for $s(0)^2 = 0.125$. If the cosine is written as a sum of two
exponentials sums of the type $\sum^\infty_{n=1} q^n/n = - \ln (1-q)$
appear in the exponent on the rhs of Eq. (45). This yields
\begin{equation}
<\tilde \psi_+^\dagger (u)\tilde \psi_+(v)> = 
\frac{e^{-i\tilde N_+(u-v)}}{1-e^{i(u-v+i0)}}
\left[\frac{(1-e^{-2/n_{c}})^2} {(1-e^{i(u-v+2i/n_{c})})(1-e^{-i(u-v-2i/n
_{c})})} \right]^{{s^2}(0)}.
\end{equation}
The calculation of $<\hat n_{m,+}>$ simplifies if we go to the
thermodynamic limit $L \to \infty, \tilde N_+/L = {\rm const. }$ Then
it is appropriate to switch to the ''physical'' field operators
$\psi_\alpha (x)$ defined as in Eq. (2) but with $c_n$ replaced by
$c_{n,\alpha}$. They are related to the auxiliary field operators
defined in (B.6) by
\begin{equation}
\psi_\alpha (x) = \frac{1}{\sqrt{L}} \tilde \psi_\alpha \left( \frac{2 \pi
  x} {L}\right).
\end{equation}
The momentum distribution $<\hat n_{k_{m},+}>$ follows from the
inversion formula (B.22) and the fact that $<\tilde \psi^\dagger_+ (u)
\tilde \psi_+ (v)>$ is a function of $u-v$ only as 
\begin{equation}
<\hat n_{k_{n},+}> = \int^{L/2}_{-L/2} dx
e^{ik_nx}<\psi^\dagger_+(x)\psi_+ (0)>.
\end{equation}
If one now takes the limit $L\to \infty, 2\pi \tilde N_+/L \equiv k_F$
one obtains for fixed $x$ from Eq. (47)
\begin{equation}
<\psi_+^\dagger (x) \psi_+(0)> = \frac{i}{2\pi} \frac{e^{-ik_Fx}} {(x +
  i0)} \left( \frac{(2/k_c)^2} {x^2 + (2/k_c)^2}\right)^{{s^2}(0)}.
\end{equation}
In contrast to the noninteracting case this 
zero temperature time-independent ''Green's
function'' for large $x$ does not decay as $(1/x)^d$, where $d = 1$ is
the spatial dimension, but as $(1/x)^{1+2s^2(0)}$. Therefore the
quantity
\begin{equation}
\alpha \equiv 2s^2(0)
\end{equation}
is called the ''anomalous dimension''. The momentum distribution in
the thermodynamic limit will be denoted $n_\alpha (k)$. The behaviour
of $n_+(k)$ near the right Fermi point is most easily studied by
looking at its derivative
\begin{equation}
\frac{dn_+(k)}{d\tilde k} = - \frac{1}{2\pi} \int^\infty_{-\infty}\left(
\frac{(2/k_c)^2}{x^2 + (2/k_c)^2}\right)^{{s^2}(0)} e^{i\tilde k x}dx,
\end{equation}
where $\tilde k \equiv k-k_F$. In the noninteracting case $s^2(0) = 0$
and $(dn_+(k)/dk)^{(0)} = -\delta(\tilde k)$, where the negative delta
function is due to the step at $k = k_F$. For finite values of
$s^2(0)$ this singularity for $\tilde k = 0$ is weakened. For $2s^2(0)
> 1$ the function of $x$ Fourier transformed in Eq. (52) decays
faster than $1/|x|$ and $(dn_+/dk)|_{\tilde k = 0}$ is {\em
  finite}. For $0< 2s^2(0)< 1$ one obtains the behaviour mentioned in
section II. As $dn_+/dk$ is an even function of $\tilde k$, it is
sufficient to consider $\tilde k > 0$. With the substitution $u = \tilde
 k x$ one obtains
\begin{equation}
\frac{dn_+}{d\tilde k} = (2\tilde k/k_c)^{\alpha-1}
(-\frac{1}{\pi})\int^\infty_0 \left(\frac{1}{u^2+(2\tilde
    k/k_c)^2}\right)^{{s^2}(0)} \cos u\; du .
\end{equation}
For $0<2s^2(0)<1$ the limit $\tilde k \to 0$ in the integrand can be
performed and the integral is finite, but $dn_+/d \tilde k$ diverges
at $\tilde k = 0$ because of the prefactor of the
integral. Integration with respect to $\tilde k$ finally yields
Eq. (16). At finite temperatures the derivative of $n_+$  at
$k_F$ diverges like $T^{\alpha -1}$ for $T \to 0$.
The special case $\alpha = 1$ requires a separate treatment
not given here.\\[0.5cm]
The power law behaviour Eq. (16) of $n_+(k)$ has been derived here for
the special potential shown in Fig.$\;$1 leading to the exponential decay
Eq. (45) of $s_n^2$. Different potentials with a different decay of
$s^2(k_n)$ lead for $|\tilde k|/k_c \ll 1$ to the {\em same} qualitative
behaviour as for our special potential if $v(k)$ goes to constant
value for $k\to 0$. This implies a constant value $s^2(0)$ and the
anomalous dimension is given by Eq. (51). The anomalous dimension can
be expressed in terms of the velocities $v_N$ and $v_J$ using
Eqs. (31) and (37) 
\begin{equation}
2s^2(0) = \frac{(v_J + v_N)/2}{\sqrt{v_N v_J}} - 1 \equiv \frac{1}{2}(K
+ \frac{1}{K} - 2),
\end{equation}
where $K \equiv \sqrt{v_J/v_N}> 0$ is less than one for repulsive
interactions and larger than one for attractive interactions.\\[0.5cm]
The power law behaviour of $n_+(k)$ for interactions $v(k)$ with a
finite limit $v(0)$ in one spatial dimension is different from the
''Fermi liquid'' type behaviour of a usual three dimensional
conductors in which the momentum distribution of the interacting
ground states has a finite discontinuity of size $0 < Z < 1$ at the
Fermi momentum \cite{Ma}. We note that for a {\em finite} one-dimensional
system and small values of $\alpha$ the distinction between the
power-law behaviour and a finite discontinuity is not
straightforward. In Fig. 4 we show the momentum distribution for a
step potential (see Fig. 1) with $v(0)/(2\pi v_F) = 1.5$ and 0.3
corresponding to $\alpha = 0.25$ and $\alpha = 0.0277$ for different
system sizes.
\begin{figure} [hbt]
\hspace{2.5cm}
\vspace{1.0cm}
\epsfig{file=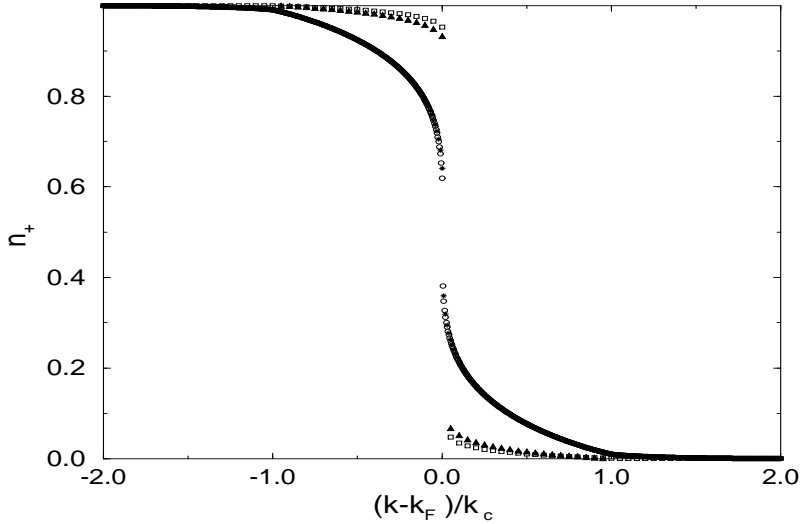,width=8cm,height=12cm,angle=-90}

\caption{ Momentum distribution for the step potential of Fig.1 
 with $v(0)/(\pi v_F)=3$ and $0.6 $
corresponding to values of the anomalous dimension
$\alpha =0.25$ and $ 0.0554$. For the larger
interaction the distribution is shown for different system sizes
(stars: $n_c=100$, circles: $n_c=200$). For the smaller interaction
the exact result for $n_c=20$ (squares) is compared to the result
from perturbation theory (triangles).   }
\end{figure}
 The size of the system is specified by the dimensionless
ratio $n_c = (k_cL)/2\pi \equiv (L/R)/2\pi$ where $R$ is the spatial
range of the interaction. For the smaller interaction we also compare
the exact result with the lowest order perturbation theory result
Eq. (15) in the range $0 < k-k_F < k_c$. As the dimensionless coupling
constant $v(0)/(2\pi v_F) = 0.3$ is not really much smaller than one
the deviations are already well visible.\\[0.5cm]
Power laws occur for the TL-model also in other physical
quantities. In the next section we present a short discussion of
one-particle spectral functions which largely determine the current in
a photoemission experiment \cite{Ma,Da}
.

\section{Spectral properties and photoemission}

For noninteracting electrons the ''density of one-particle states per
unit volume'' (DOS) $\rho_0 (\epsilon)$ plays an important role for
the calculation of thermodynamic properties \cite{AM}. In one dimension it
is defined as
\begin{equation}
\rho_0(\epsilon)\equiv \frac{1}{L} \sum_{k_n} \delta (\epsilon -
\epsilon_{k_n }).
\end{equation}
In the thermodynamic limit $\sum_k(\cdot) \to \frac{L}{2\pi}\int (\cdot)dk$
and for a system of noninteracting right movers $(\epsilon_k = v_Fk)$
one obtains $\rho_0(\epsilon) = 1/(2\pi v_F)$, i.e. an energy
independent DOS. For photoemission only the occupied states matter and
one defines the ''occupied DOS'' $\rho_0^<(\epsilon)$ by multiplying
the delta function in Eq. (55) with the Fermi function
$f(\epsilon_{k_n})$, which at $T = 0$ is the step function at the
Fermi energy $\epsilon_F$. This simple DOS concept has to be
generalized properly when the fermions interact. For a translational
invariant system as discussed in the previous sections one defines
\begin{equation}
\rho^< (\epsilon) \equiv \frac{1}{L} \sum_n <\phi_0(N)|c^\dagger_n
\delta(\epsilon + H - E_0(N))c_n| \phi_0(N)>
\end{equation}
as the generalization of the occupied DOS relevant for photoemission
\cite{Ma}. In the {\em noninteracting} limit $c_n|\phi_0(N)>$ is the
normalized Slater determinant where the state with momentum $k_n$ has
been removed relative to the N-electron ground state. As its energy is
given by $E_0(N)- \epsilon_{k_n}$ the spectral function
$\rho^<(\epsilon)$ reduces to the sum of delta functions as in
Eq. (55) restricted to the occupied momenta. The individual terms in
the sum of the rhs of Eq. (56) enter ''angular resolved
photoemission''\cite{MS,V}, while $\rho^<(\epsilon)$ is needed for the
description of angular integrated photoemission \cite{Da}. For translational
invariant systems expectation values with $c^\dagger_n$ replaced by
$c^\dagger_{n'}$ with $n' \neq n$ in Eq. (56) vanish because of
momentum conservation and one can write $\rho^<(\epsilon)$ in the form
\begin{eqnarray}
\rho^<(\epsilon) 
&=&
 <\phi_0(N)|\psi^\dagger(x=0)\left(\frac{1}{2\pi} \int
e^{i(\epsilon + H-E_0(N))t}dt\right) \psi (x = 0)|\phi_0(N)> \\
&=&
\int <\phi_0(N)|\psi^\dagger(x=0)\psi(x=0,t)|\phi_0(N)>e^{i\epsilon t}
\frac{dt}{2\pi} 
 \equiv \int iG^<(t)e^{i\epsilon t}\frac{dt}{2\pi},\nonumber 
\end{eqnarray}
where $\psi(x,t) = e^{iHt} \psi(x)e^{-iHt}$ is the field operator in
the Heisenberg representation. If we switch from the original
nonrelativistic model to the TL-model the spectral function acquires an
additional $\alpha$ label. In the following we sketch the calculation
of $\rho^<_+(\epsilon)$. For that purpose it is convenient to express
the field operator $\tilde \psi_+(v)$ Eq. (B.8) in terms of the
$\alpha_n,\alpha^\dagger_n$ instead of the $b_n, b^\dagger_n$ because the former have a simple time
dependence. In $b_m = c_m \alpha_m + s_m \alpha^\dagger_{-m}$ the
operators $\alpha_m$ and $\alpha^\dagger_{-m}$ commute. Therefore
$e^{i\phi_+(v)}$ (and $e^{i\phi_+^\dagger(v)}$) in Eq. (B.8) can be written
as a product of two exponentials with the annihilation operators to the
right. After using the Baker-Hausdorff formula once in order to
complete the process of ''normal ordering'' (bringing all annihilation
operators to the right of the creation operators) one obtains
\begin{eqnarray}
\psi_+ (x) & = & 
\frac{A(n_c)}{\sqrt{L}}\hat O_+\left(\frac{2\pi x}{L}\right)
e^{i\chi^\dagger_+(x)} e^{i\chi_+(x)} \\ \nonumber
{\rm with}\quad i\chi_+(x)& = &
\sum^\infty_{m=1}\frac{1}{\sqrt{m}}\left(c_m e^{ik_mx} \alpha_m - s_m
  e^{-ik_mx} \alpha_{-m}\right)
\end{eqnarray}
and $A(n_c) \equiv e^{-\sum^{\infty}_{n=1} s_n^2/n}$.\\[0.3cm]

This is one of the most important formulas for the physics of
one-dimensional interacting fermions. The time dependent operator
$\psi_+(x,t)$ follows from Eq. (58) by replacing $\alpha_m$ and
$\alpha_{-m}$ by $\alpha_m e^{-i\omega_mt}$ and
$\alpha_{-m}e^{-i\omega_mt}$ and $U_+$ in $\hat O_+$ (see Eq. (B.23)) by
$U_+(t)$. Various kinds of time dependent correlation functions can
quite simply be calculated using this result \cite{LP,H,MS,V,SM2}. Here
we restrict ourselves to $G_+^<(t)$ in order to obtain
$\rho_+^<(\epsilon)$.\\[0.5cm]
As $U_+$ commutes with the bosonic operators we first show that the
particle number changing operators $U^\dagger_+$ and $U_+(t)$ lead to
a simple time dependent factor. Using (B.17) one obtains
\begin{equation}
U^\dagger_+ e^{iHt}U_+ e^{-iHt}|\{0\}_\alpha, \tilde N_+, \tilde N_->
= e^{-i(E_0(\tilde N_+,\tilde N_-) - E_0(\tilde N_+ - 1, \tilde
  N))t}|\{0\}_\alpha, \tilde N_+, \tilde N_->.
\end{equation}
As $\psi_+(x)$ in Eq.(58) is normal ordered in the $\alpha$'s one has
to use the Baker-Hausdorff formula only once to normal order
$\psi^\dagger_+(x=0) \psi_+(x=0,t)$. This yields
\begin{eqnarray}
ie^{i\mu t}G^<_+(t) 
&=&
\frac{A^2(n_c)}{L} e^{[\chi_+(x=0,t=0),\chi_+^\dagger (x=0,t)]}\\ \nonumber
&=&
\frac{A^2(n_c)}{L} \exp\left\{ \sum^\infty_{n=1} \frac{1+2s^2_n}{n}
  e^{i\omega_nt}\right\} 
\end{eqnarray}
where $\mu \equiv E_0(\tilde N_+, \tilde N_-) - E_0(\tilde N_+ -1,
\tilde N_-)$ is the chemical potential, which we ``eliminate'' by
discussing $\rho^<_+$ as a function of $\tilde \epsilon \equiv
\epsilon - \mu$. The spectral function can be calculated analytically
in the ''universal'' {\em low energy regime} $-k_c{\rm min}\; (v_F,v_c) \le
\tilde \epsilon \le 0$ for the step potential which has a strictly
linear boson dispersion $\omega_n = v_c|k_n|$ for $|k_n| \le k_c$ (see
Fig. 2) and a different linear dispersion $\omega_n = v_F|k_n|$ for
$|k_n| > k_c$. This implies that we can write $G^<_+(t)$ as a product of
two power series in $\exp(i2\pi v_Ft/L)$ and $\exp(i 2\pi v_ct/L)$
\begin{equation}
iL e^{i\mu t}G^<_+(t) = A^2(n_c)\left( 1 + \sum^\infty_{m=n_c+1}
  a_m^{(n_c)}
  e^{im\left(\frac{2\pi}{L}v_Ft\right)}\right)\left(\sum^\infty_{n=0} 
d_n^{(n_c)} e^{in\left( \frac{2\pi}{L}v_c t\right)}\right).
\end{equation}
Now the Fourier integral can be performed analytically. For $v_Fk_c
\le \tilde{\epsilon} \le 0$ the $a_m^{(n_c)}$  and for $v_ck_c \le
\tilde{\epsilon} \le 0 $ the $d_m^{(n)}$ with $m > n_c$ do not
contribute and one obtains in the universal regime $-k_c \min (v_F, v_c)
\le \tilde{\epsilon} \le 0$
\begin{equation}
L \rho_+^< (\tilde{\epsilon}) = A^2(n_c) \cdot \sum_{n \ge
  0}~d_n^{(n_c)} \delta \left( \tilde{\epsilon} + n~\frac{2 \pi}{L}~v_c \right)~.
\end{equation} 
The $d_n^{(n_c)}$ for $n \le n_c$ can be given analytically as the sum
in the exponent in Eq. (60) is of the form $(1 + \alpha) \cdot
\sum^{n_c}_{n = 1} z^n/n$ with $z = \exp (i 2 \pi v_ct/L)$ and $\alpha =
2 s^2 (0)$. For the calculation of the $d_n^{(n_c)}$ for $n \le n_c$
one makes no error to extend the sum to infinity, which yields $\exp
\left[ (1 + \alpha)~\sum_n z^n/n \right] = (1 - z)^{-(1 +
  \alpha)}$. Now one can use the well known power expansion of this
function. Alternatively one can show that the recursion relation in
(A4) of I can simply be solved explicitly. If we also use $\sum^{n_c}_{n
  = 1}1/n \rightarrow \ln n_c + C$, where $C$ = 0.57721
... is Euler's constant \cite{AS} we find the {\em exact} weights of the
delta peaks in Eq. (62)
\begin{equation}
A^2(n_c) d_n^{(n_c)} =~\frac{e^{-\alpha C}}{n_c^\alpha}~\prod_{\ell =
  1}^n \left( 1 + \frac{\alpha}{\ell} \right)
\end{equation} 
for $n_c \ge n \ge 1$ and $d_0^{(n_c)} = 1$. For $n_c \ge n \gg 1$ one
can use the product representation of Euler's Gamma function \cite{AS} to
write the weights as 
\begin{equation}
A^2(n_c) d_n^{(n_c)} = ~\frac{e^{-\alpha C}}{\Gamma(1 +
    \alpha)}~\left( \frac{n}{n_c} \right)^\alpha \left( 1 + O (1/n_c) 
\right)~.
\end{equation}
Now the thermodynamic limit can be performed in Eq. (62) and one
obtains with $\epsilon_c \equiv v_c k_c$ for $\epsilon \le \mu$ in the
universal regime 
\begin{equation}
\rho_+^< (\epsilon) =~\frac{e^{-\alpha C}}{\Gamma(1 +
    \alpha)}\cdot 
\frac{1}{2 \pi v_c}~\left( \frac{\mu -
    \epsilon}{\epsilon_c} \right)^\alpha~.
\end{equation}
The spectral weight near (below) the chemical potential $\mu$ is
suppressed by the same power law behaviour as in $n_+ (k)
- 1/2$. This type of power law occurs also for potentials with a finite
$v(0)$ but a different decay for momenta large compared to
$k_c$. But then the prefactor is different and the power law behaviour
does not hold {\em exactly} in a finite range as Eq. (65) but
only {\em asymptotically} in the limit $(- \tilde{\epsilon})/\epsilon_c
\rightarrow 0$ \cite{M}. This power law suppression of spectral weight was
claimed to be experimentally verified in photoemission experiments with
an organic compound which consists of weakly coupled chains \cite{Da}.

Reliable results for more complicated models of interacting fermions
can often only be obtained using numerical methods for systems of
rather small size. As our result Eq. (63) is exact also for finite
systems we shortly discuss how one can extract the anomalous
dimension even for small values of $n_c$. As Eq. (64) only holds for
$n \gg 1$ it is {\em not} useful to fit the weights to the power
law form. It is much easier to look at the scaling behaviour of the
peak at $\epsilon = \mu$ with weight $A^2(n_c) d_0^{(n_c)} = e^{-
  \alpha C}/(n_c)^\alpha$ with system size. From a log-log-plot one can
directly infer the value of $\alpha$.

For $- \tilde{\epsilon} \gg \epsilon_c$ the spectral function
$\rho_+^< (\tilde{\epsilon})$ approaches the noninteracting value
$1/(2 \pi v_F)$. The crossover from the power law behaviour to this
high energy limit is discussed in \cite{SM2}.

Finally we want to point out that the low energy power law behaviour
of $\rho_+^< (\epsilon)$ given by Eq. (65) holds for
{\em arbitrary} values of $\alpha$. {\em No} additional
linear contribution occurs for $\alpha > 1$ as in the case of the
momentum distribution $n_+ (k) - 1/2$. This difference is easily
lost without a proper treatment of the cut-offs.

\section{Spin-charge separation in the model including spin}
Electrons are spin one-half particles and for their description it is
 necessary to include the spin degree of freedom in the model. We
 choose a fixed quantization axis and denote the two spin states by
 $\sigma = \uparrow,\downarrow$. Then the field operator $\psi(x)$
 acquires a spin index as the $c_n$. The density operator of spin
 projection $\sigma$ is defined as $\rho_\sigma(x) \equiv 
\psi_\sigma^\dagger(x)\psi_\sigma(x)$ and in the two-body interaction 
in Eq. (6)
 $V(x-x')\hat{\rho}(x) \rho(\hat{x})$ is replaced by $\sum_{\sigma,
 \sigma'}~V_ {\sigma \sigma'}(x - x') \hat{\rho}_\sigma(x)
 \hat{\rho}_{\sigma'} (x')$. Usually the interaction $V_{\sigma
 \sigma'}(x-x')$ is independent of the spin indices \cite{Ma}. After
 the linearization and introduction of the right and left movers the
 $\hat \rho_{n,\alpha}$ in Eq. (20) obtain an additional spin index, as
 well as the boson operators in Eq. (27). The TL-Hamiltonian for spin 1/2
  then reads 
\begin{eqnarray}
& & H_{TL}^{(1/2)} =~\sum_{\sigma, \sigma'}~ \left\{ \sum_{n>0} \left[ k_n
  \left( v_F \delta_{\sigma \sigma'} + ~\frac{g_4^{\sigma
  \sigma'}(k_n)}{2 \pi} \right)  \left( b^\dagger_{n,\sigma} b_{n, \sigma'}
  + b^\dagger_{-n, \sigma} b_{-n, \sigma'} \right) \right. \right. \nonumber
  \\
& & \hspace*{2cm} \left. + ~\frac{k_n g_2^{\sigma
  \sigma'}(k_n)}{2 \pi} \left( b^\dagger_{n,\sigma} b^\dagger_{-n, \sigma'} +
  b_{-n, \sigma'} b_{n \sigma} \right)  \right]\\    
& & \hspace*{1.5cm} \left. + ~\frac{\pi}{L}~ 
\left[ \left (  v_F \delta_{\sigma \sigma'}+ \frac{ g_4^{\sigma \sigma'}
  (0)}{2\pi}\right ) ~\sum_\alpha {{\cal N}}_{\alpha, \sigma} {{\cal
  N}}_{\alpha, \sigma'} + \frac{ g_2^{\sigma \sigma'}(0)}{\pi} {{\cal
  N}}_{+,\sigma} {{\cal N}}_{-,\sigma'} \right] \right\}~.\nonumber
\end{eqnarray}
The interaction matrix elements $g_\nu^{\sigma \sigma'}(k)$ are taken
as depending on the relative spin only, i.e. $g_\nu^{\sigma
  \sigma}$ and $g_\nu^{\sigma, -\sigma}$ are independent of the spin
label $\sigma$. The interaction term in Eq. (66) couples the electrons
with different spin. With the assumption on the $g_\nu^{\sigma
  \sigma'}$  it
is possible to switch to new boson operators $b_{n,a}$ with $a = c, s$
\begin{eqnarray}
b_{n,c} & \equiv &\frac{1}{\sqrt{2}}~ (b_{n,\uparrow} + b_{n,\downarrow})
\nonumber \\
b_{n,s} & \equiv & \frac{1}{\sqrt{2}}~ (b_{n \uparrow} - b_{n, \downarrow})~~,
\end{eqnarray}
which obey $\left[ b_{a,n}, b_{a', n'} \right] = 0~;~ \left[
  b_{a,n}, b_{a', n'}^\dagger  \right] =  \delta_{a a'} \delta_{n n'}
  {\hat 1}$. The kinetic energy can also be expressed in terms of
  ``charge'' ($c$) and ``spin'' ($s$) Bose operators using
  $b^\dagger_{n,\uparrow} b_{n,\uparrow} + b^\dagger_{n \downarrow}
  b_{n \downarrow} = b_{n,c}^\dagger b_{n,c} + b_{n,s}^\dagger
  b_{n,s}$. If we define the interaction matrix elements
  $g_{\nu,a}(q)$ via
\begin{eqnarray} 
g_{\nu,c}(q) & \equiv & g_\nu^{\sigma \sigma}(q) + g_\nu^{\sigma,
  -\sigma}(q) \nonumber \\
g_{\nu,s}(q) & \equiv & g_\nu^{\sigma \sigma}(q) - g_\nu^{\sigma, -\sigma}(q)~,
\end{eqnarray}
and define ${\cal N}_{c(s)}\equiv ({\cal N}_{\uparrow} \pm {\cal N}_{
\downarrow})/\sqrt 2
$ we can write $H^{(1/2)}_{TL}$ as
\begin{equation}
H^{(1/2)}_{TL} = H_{TL,c} + H_{TL,s}~,
\end{equation}
where the $H_{TL,a}$ are of the form Eq. (30,31) but the interaction
matrix elements have the additional label $a$. The two terms on the
rhs of Eq. (69) {\em commute}, i.e. the ``charge'' and ``spin''
excitation are completely independent. This is usually called
``spin-charge separation'' \cite{V2}. The ``diagonalization'' of the two
separate parts proceeds exactly as in section III and the low energy
excitations are ``massless bosons'' $\omega_{n,a} = v_a |k_n|$ with
the {\em charge velocity} $v_c$ and the {\em spin
  velocity} $v_s$ usually different. For a spin independent
interaction $g_{\nu,s}$ in Eq. (68) vanishes and $v_s$ equals the
Fermi velocity $v_F$. As discussed in the next section the Hamiltonian
$H_{TL}^{(1/2)}$ with properly chosen interaction matrix elements
describes correctly the low energy physics of more complicated $1d$
models e.g. on a lattice. Then $v_s$ and $v_F$ are generally different even
if the interaction on the lattice is spin independent.

The bosonization of the field operator $\psi_{\alpha, \sigma}(x)$
proceeds exactly as in appendix B. In order to calculate quantities
like the momentum distribution $\langle \hat n_{k, +, \sigma} \rangle$
it is then useful to express the $b_{n,\sigma}$ in $i \phi_{+,\sigma}(v)$
like in (B.9) in terms of the charge and spin bosons
\begin{equation}
i \phi_{+, \sigma}(v) =
~\frac{1}{\sqrt{2}}~\sum^\infty_{n=1}~\frac{e^{i n v}}{\sqrt{n}}
(b_{n,c} \pm b_{n,s}) \equiv i\phi_{+,c}(v) \pm
i\phi_{+,s}(v)~,
\end{equation}
where the plus (minus) sign holds for the up (down) spin.
Therefore e.g. the function  $\langle \psi^\dagger_{+,\uparrow}(x)
\psi_{+,\uparrow}(0) \rangle$ is a {\em product} of a charge and
a spin part. The separate factors are calculated as in the spinless
case. The only difference is a factor $1/2$ in the exponent due to the
additional factor $1/\sqrt{2}$ in Eq. (70). The final result is
exactly of the form Eq. (45) but with $2 s^2_n$ replaced by $s_{n,c}^2
+ s_{n,s}^2$. This leads to the power law behaviour
presented in Eq. (16) as in the
spinless case but the anomalous dimension given by
\begin{equation}
\alpha = s^2_c(0) + s^2_s(0) \equiv \alpha_c + \alpha_s~.
\end{equation}
The individual contributions can be expressed in terms of the $K_a
\equiv \left( v_{J,a}/v_{N,a} \right)^{1/2}$ as $\alpha_a = (K_a -1)^2
/(4 K_a)$ similar to Eq. (54). Also the result for $G_{+,\sigma}^<(t)$
needed in the calculation of the integrated spectral function
$\rho^<_{+,\sigma}(\epsilon)$ takes the form of a product of a
``spin'' and a ``charge'' factor. In each factor the weight $1 +2
s_n^2$ in the exponent of Eq. (60) is replaced by $1/2 +
s^2_{n,a}$. Therefore the Fourier transform of the
{\em individual factors} are in the low energy regime of the
power law form Eq. (65) with $\alpha$ replaced by $-\frac{1}{2} +
\alpha_a$. The convolution of the two power laws finally for
$\epsilon \le \mu$ yields $\rho_{+,\sigma}^< (\epsilon) \sim (\mu -
\epsilon)^\alpha$ in the asymptotic regime as in the spinless
model. Alternatively this follows more directly from the fact
that  $G_{+,\sigma}^<(t)$ decays as $t^{-(1+\alpha)}$ for large times.

 In order to see a drastic difference between the model with and
without spin, one has to calculate the individual terms in the sum in
Eq. (56). The delta peaks of the noninteracting model are broadened
into {\em one} ``infrared'' power law threshold in the model
without spin \cite{LP,MS,SM2} and {\em two} power law singularities
in the model including spin \cite{MS,V,SM2}.

\section{The Luttinger liquid phenomenology}
We have seen that the low energy physics of the $TL$-model is determined
by two parameters $v_c$ and $K$ or $v_N$ and $v_J$ in the spinless
model and four parameters $(v_c, K_c; v_s, K_s)$ or $\left( v_{N_c},
  v_{J_c}, v_{N_s}, v_{J_s} \right)$ in the model including spin. They
also determine the low temperature thermodynamics of the model. We
shortly discuss the model including spin. The low temperature specific
heat of a linear chain is given by $\gamma T$, where the constant
$\gamma$ is inversely proportional to the sound velocity. This is the
$1 d$ version of Debye's law \cite{AM,SM}. The ratio of $\gamma$ and
the corresponding value $\gamma_0$ for the noninteracting fermions is therefore
\begin{equation}
\gamma/\gamma_0 =~\frac{1}{2}~\left( \frac{v_F}{v_c}~+~\frac{v_F}{v_s}
\right)~.
\end{equation}
At zero temperature we can express the pressure $p$ as $p = -(\partial
E_0/\partial L)_N$ and the compressibility $\kappa \equiv -(\partial
L/\partial p)_N/L = \left( L(\partial^2 E_0/\partial L^2)_N \right)^{-1} =
\left( (N^2/L)(\partial^2 E_0/\partial N^2)_L \right)^{-1}$. Because
of ${{\cal N}}_c \equiv \left( {{\cal N}}_\uparrow +
  {{\cal N}}_\downarrow \right) /\sqrt 2$ the $N$-dependence of the
ground state energy of the $TL$-model using Eq. (30) is given by
$v_{N_c}(\pi/4 L)N^2$ leading to 
\begin{equation}
L\left( \frac{\partial^2 E_0}{\partial N^2} \right)_L
=~\frac{\pi}{2}~v_{N_c} =~\frac{\pi}{2}~\frac{v_c}{K_c}.
\end{equation}
The compressibility ratio is therefore given by
\begin{equation}
\kappa/\kappa_0 =~\frac{v_F}{v_c}~K_c~.
\end{equation}
The factor $1/2$ in Eq. (73) is absent in the spinless case.
We leave it as an exercise to show that one obtains similarly for the
spin susceptibility
\begin{equation}
\chi / \chi_0 =~\frac{v_F}{v_s} \cdot K_s~.
\end{equation}
These relations show that the $K_a$ which determine the contributions
$\alpha_a$ to the anomalous dimension $\alpha$
and therefore the power law behaviour of the Green's functions
 also enter the low
temperature thermodynamics.

 It was an important insight by Haldane \cite{H}
to realize that the low energy physics of the $TL$-model is {\em
  generic} for interacting fermions in one dimension. In the language
of the renormalization group \cite{V2} the TL-Hamiltonian is the {\em fixed
point} Hamiltonian for 1d fermions with a repulsive interaction.
This is the main reason for the importance of the TL-model for
the physics in one dimension.

Usually the LL parameters $K_a,v_a$ of the effective TL-model which describes
the low energy physics of a given model can only be calculated
approximately. An example is the original Tomonaga model with
the quadratic energy dispersion (with or without spin) where it
is  {\em not} assumed that the interaction cutoff is 
much smaller than the Fermi momentum,
 i.e. an {\em arbitrary } repulsive interaction is used.
An exception is the extreme short range limit $V(x)=V_0\delta(x)$.
For the spinless model $\hat V$ then vanishes because the Pauli
principle forbids two fermions at the same space point
($\psi(x)\psi(x)=0$ in Eq. (1)) . For the corresponding model including
spin ground state properties can be calculated exactly using the
Bethe-ansatz method \cite{V2}. The ratios $K_c/v_c$ and $K_s/v_s$
of the effective TL-model which describes the low energy physics of 
the model can be inferred by calculating the compressibility and 
spin susceptibility using Eqs. (74) and (75). The charge and spin
velocities can be obtained as $(E^{c,(s)}_0(2\pi/L)-E_0)/(2\pi/L)$,
where $E^{c(s)}_0(2\pi/L)$ is the lowest energy for charge (spin)
excited states with total momentum $2\pi/L$. From this information
one can obtain the anomalous dimension, which determines the 
decay of the Green's function without the need to calculate this
function. For 1d models on a lattice, which cannot be solved exactly
by the Bethe-ansatz method one can determine the LL parameters
approximately by numerically calculating the quantities discussed
above for small systems. In the case of weak interactions perturbation
theory can be used. We show this to leading order for the spinless
Tomonaga model with a short range interaction.
The ground state energy up to first order in the interaction is given
by the expectation value of $H$ in the Fermi sea $|F>$
\begin{equation}
E^{(1)}_0=~\sum^{n_F}_{n=-n_F}\frac{1}{2m_e}(\frac{2\pi n}{L})^2
+\frac{1}{2L}\sum^{n_F}_{n=-n_F}\sum^{n_F}_{n'=-n_F}(v(0)-v(k_n-k_n')).
\end{equation} 
The interaction contributes the Hartree and the Fock term. In the
large system limit the sums can be replaced by integrals in the
standard way and one obtains
\begin{equation}
L\left( \frac{\partial^2 E^{(1)}_0}{\partial N^2} \right)_L 
=\pi v_F +v(0)-v(2k_F) \equiv \pi (v_c/K)^{(1)}.
\end{equation}
In order to obtain the charge velocity to linear order in the
interaction we have to calculate the expectation value of $H$ in
the state $|\frac{2\pi}{L}>=c^{\dagger}_{n_F+1}c_{n_F}|F> $.
Taking the difference with $E_0$ and dividing by $2\pi/L$ yields
\begin{equation} 
(v_c)^{(1)}=v_F +(v(0)-v(2k_F))/(2\pi).
\end{equation}
The contribution of the interaction is due to the different Fock
energies in the states $|\frac{2\pi}{L}>$ and $|F>$.
Combining the results from Eqs. (77) and (78) gives $K=
1-(v(0)-v(2k_F))/(2\pi v_F)+...$ which leads to
\begin{equation}
\alpha^{(2)}=\frac{1}{2} \left (\frac{v(0)-v(2k_F)}{2\pi v_F}\right )^2.
\end{equation} 
If one generalizes the calculation of $<\hat n_m>^{(2)}$
in section II to the case of a short range interaction one obtains the
same result for the anomalous dimension. This shows to 
leading order perturbation theory that the low energy physcis of the 
Tomonaga model with an arbitrary range interaction is that of the TL
model
with $g_2(0)=g_4(0)=v(0)-v(2k_F)$.\\

Because of the limited length of the paper the rapidly expanding field
of electronic transport in one-dimensional interacting Fermi systems
was not discussed. Important insights can rather quickly be obtained
from the simple exercise 
of normal ordering of $\psi_+^{\dagger }(x)  \psi_-(x)$ with the use
of Eq.(58) and the corresponding expression for the left movers \cite{Mattis}.
The interested reader should look up the relevant references in the
review article cited \cite{V2},
 especially the important work by Kane and Fisher
\cite{KF}.

\section*{Acknowledgments}
I would like to thank L.Bartosch, P.Kopietz and V.Meden for a critical
reading of the manuscript and the latter also for the pleasant
collaboration on several issues discussed in this paper.
The now more extended discussion of commutation relation of the phase
operator and the particle number operator was stimulated by a recent
preprint (cond-mat/9805275) by J.von Delft and H.Schoeller.

 \newpage 
\appendix

\renewcommand {\theequation}{\thesection.\arabic{equation}}
\section{The Kronig identity}
\setcounter{equation}{0}
In this appendix we generalize an identity found by Kronig
in the context of the (now long forgotten) ``neutrino theory of
light''
\cite{Kro} to the case of a finite cut-off $k_B$.
For the right movers it reads 
\begin{equation}
\sum^\infty_{l = 1}~l b_l^\dagger b_l =~\sum^\infty_{m = -m_0} (m +
m_0) c_{m,+}^\dagger c_{m,+} - ~\frac{1}{2}~\left(
  {{\cal N}}_+^2 - {{\cal N}}_+ \right)~.
\end{equation}
For the proof we simply evaluate the lhs using Eqs. (20) and (27)
\begin{eqnarray}
\sum^\infty_{l = 1}~l b_l^\dagger b_l & = & \sum^\infty_{l =
  1}~\sum^\infty_{n = -m_0}~c^\dagger_{n + l,+} c_{n,+}
  c^\dagger_{n,+} c_{n + l,+}~+~\sum^\infty_{l = 1}~\sum_{n \not= n'}~
  c^\dagger_{n + l,+} c_{n, +} c^\dagger_{n',+} c_{n'+l,+} \nonumber \\
& \equiv & \hat{A}_1 + \hat{A}_2~,
\end{eqnarray}
where we have split the double sum over $n$ and $n'$ in the
contributions from $n = n'$ and $n \not= n'$. We will show that
 the operator $\hat{A}_1$
yields the rhs of Eq. (A.1) while $\hat{A}_2$ vanishes
\begin{eqnarray}
\hat{A}_1 & = & \sum^\infty_{l = 1}~\sum^\infty_{n  = -m_0}~\hat{n}_{n + l,
  +} ~- ~\sum^\infty_{l = 1}~\sum^\infty_{n = -m_0}~\hat{n}_{n + l,+}
  \hat{n}_{n,+}\\ \nonumber
& = & \sum^\infty_{m = -m_0}~(m + m_0) \hat{n}_{m,+} -~\frac{1}{2}
  \left( {{\cal N}}^2_+ - {{\cal N}}_+ \right)~.
\end{eqnarray}
The second equality follows from a change of summation variables and
using $\hat{n}_{m,+}^2 = \hat{n}_{m,+}$. For the discussion of
$\hat{A}_2$ we split the double sum into the contributions from $n >
n'$ and $n < n'$
\begin{equation}
\hat{A}_2 = ~\sum^\infty_{l = 1}~ \left\{ \sum_{n > n'}~c^\dagger_{n +
    l,+} c_{n,+} c^\dagger_{n',+} c_{n'+l,+} + ~\sum_{n <
    n'}~c^\dagger_{n+l,+} c_{n,+} c^\dagger_{n',+} c_{n'+l,+} \right\}~.
\end{equation}
In the first term we write $n = n' + l'$ and use $l' \ge 1$ as the new
summation variable and in the second term we write $n'=n+l'$. This yields
\begin{equation}
\hat{A}_2 =~\sum^\infty_{l = 1}~\sum^\infty_{l'=1}~\left\{
  \sum^\infty_{n'= - m_0}~c^\dagger_{n'+l+l',+} c^\dagger_{n',+}
  c_{n'+l,+} c_{n'+l',+} + ~\sum^\infty_{n = - m_0}~c^\dagger_{n +
  l,+} c^\dagger_{n + l',+} c_{n + l + l',+} c_{n,+} \right\}~.
\end{equation}
As the products of the four operators are both
{\em antisymmetric} in $l$ and $l'$ the operator $\hat{A}_2$
vanishes after performing the summations, which completes the proof. If
we multiply Eq. (A.1) with $v_F(2 \pi/L)$ and use the analogous
relation to (A.1) for the left movers we obtain Eq. (29) for the
kinetic energy $T_{TL}$. We note that in the proof {\em no} use is made of
the commutation relations of the $b_l$ and $b_l^\dagger$. The Kronig
identity holds as an {\em exact} operator identity without
performing the limit $k_B \rightarrow \infty$.

\section{Bosonization of the field operator}
\setcounter{equation}{0}
In this appendix we generalize chapter V. of I on the bosonization of
fermionic operators to the case of a {\em single} fermion operator .
A new straightforward construction of the particle number changing
part is given which simplifies earlier presentations \cite{H}.

We start with a simple relation which holds for operators
$b,b^\dagger$ obeying $[b,b^\dagger] = \hat 1$. It reads
\begin{equation}
[b,e^{\lambda b^{\dagger}}] = \lambda e^{\lambda b^{\dagger}}~.
\end{equation}
To prove it one differentiates $b(\lambda)\equiv e^{-\lambda
  b^{\dagger}} b e^{\lambda b^{\dagger}}$ with respect to $\lambda$
  and obtains $db(\lambda)/d\lambda =  \hat 1$. Therefore
  $b(\lambda) = b + \lambda \hat 1 = e^{-\lambda b^\dagger} b e^{\lambda
  b^\dagger}$. Multiplying from the left with $e^{\lambda b^\dagger}$
  yields (B.1). Next we define the following operators linear in the
  boson operators $b_n, b_n^\dagger$
\begin{equation}
A_+ \equiv \sum_{n \neq 0} \lambda_n b^\dagger_n
\quad ; \quad B_- \equiv \sum_{n \neq 0} \mu_n b_n
\end{equation}
with arbitrary constants $\lambda_n$ and $\mu_n$. Using (B.1) and
elementary generalizations one can prove the commutation relations
\begin{eqnarray}
[b_m, e^{B_{-}} e^{A_{+}}] 
& = & 
\lambda_m e^{B_{-}} e^{A_{+}}\\ \nonumber
[b_m^\dagger, e^{B_{-}} e^{A_{+}}] 
& = & 
-\mu_m e^{B_{-}} e^{A_{+}}~.
\end{eqnarray}
In the following we consider fermionic operators $\hat S$ which obey
$[b_m,\hat S] = -\lambda_m \hat S$ and $[b_m^\dagger , \hat S] = \mu_m \hat
S$. Then the operator $\hat O\equiv \hat S e^{B_-} e^{A_+}$ commutes
with all $b_m$ and $b^\dagger_m$. We therefore write
\begin{equation}
\hat S = \hat O e^{-A_+} e^{-B_-}, 
\end{equation}
and subsequently 
construct $\hat O$ such that both sides of Eq. (B.4) yield identical
matrix elements.\\[0.5cm]
In the following we present the bosonization of the field operator for
the right movers described by the $c_{l,+}$ and just mention the
corresponding result for the left movers. The $c_{l,+}$
obey for $m > 0$ the following commutation relations
\begin{equation}
[b_m,c_{l,+}] = - \frac{1}{\sqrt{m}}c_{l+m,+}\quad , \quad
[b^\dagger _m, c_{l,+}] = - \frac{1}{\sqrt{m}}c_{l-m,+},
\end{equation}
The auxiliary field operator $\tilde \psi_+(v)$ defined
as 
\begin{equation}
\tilde \psi_+(v)\equiv \lim_{m_{0} \to \infty} \sum_{l\in I_{+}} e^{ilv}
c_{l,+} = \sum^\infty_{l = -\infty} e^{ilv}c_{l,+}.
\end{equation}
obeys commutation relations as the operator $\hat S$ discussed above
\begin{equation}
[b_m, \tilde \psi_+(v)] = - \frac{1}{\sqrt{m}}e^{-imv}\tilde
\psi_+(v)\; ; \; [b^\dagger_m, \tilde\psi_+ (v)] = - \frac{1}{\sqrt{m}}
e^{imv} \tilde \psi_+ (v)~.
\end{equation}
Therefore it is of the form
\begin{equation}
\tilde \psi_+ (v) = \hat O_+(v) e^{i\phi_+^\dagger (v)} e^{i\phi_+(v)},
\end{equation}
where the operator $i\phi_+ (v)$ is given by
\begin{equation}
i\phi_+(v) = \sum^\infty_{n = 1}\frac{e^{inv}}{\sqrt{n}} b_n~.
\end{equation}
As $\tilde \psi_+ (v)$ reduces the number of right movers by one, the
operator $\hat O_+(v)$, which commutes with all $b_m,b^\dagger_m$,
also must have this property. In order to determine $\hat O_+ (v)$ we
work with the eigenstates of the noninteracting system (compare Eq. (41))
\begin{equation}
|\{m_l\}_b, \tilde N_+, \tilde N_-> \equiv \prod_{l}
\frac{(b^\dagger_l)^{m_l}} {\sqrt{m_l!}}| \{0\}_b, \tilde N_+, \tilde N_->~.
\end{equation}
After the limit $m_0 \to \infty$ the $\tilde N_\alpha$ can run from
minus to plus infinity. It is easy to see that $\hat O_+(v)|\{0\}_b,
\tilde N_+, \tilde N_->$ has no overlap to excited states
\begin{eqnarray}
\lefteqn{<\{m_l\}_b, \tilde N_+ -1, \tilde N_-|\hat O_+(v)|\{0\}_b,
  \tilde N_+,\tilde N_->
\;  =} \hspace{1.5cm} \nonumber \\
& & < \{0\}_b, \tilde N_+-1, \tilde N_-| \prod_l
\frac{(b_l)^{m_l}}{\sqrt{m_l!}}\hat O_+(v)|\{0\}_b, \tilde N_+, \tilde N_->~.
\end{eqnarray}
As $\hat{O}(v)$ commutes with the $b_l$ the rhs of Eq. (B.11) vanishes
unless all $m_l$ are zero. This implies
\begin{equation}
\hat O_+(v)|\{0\}_b, \tilde N_+, \tilde N_-> = c^+_{\tilde N_+,\tilde
  N_-} (v)|\{0\}_b, \tilde N_+-1, \tilde N_->~,
\end{equation}
where $c^+_{\tilde N_+,\tilde
  N_-} (v)$ is a c-number.
In order to determine the $c_{\tilde{N}_+,\tilde{N}_-}^+$ we
calculate 
$\langle \{0\}_b , \tilde{N}_+ -1, \tilde{N}_- |\tilde{\psi}(v)|
\{0\}_b, \tilde{N}_+, \tilde{N}_- \rangle$ using (B.6) as well as
(B.8). In the calculation of the matrix element with the fermionic form
(B.6) we use Eq. (40) which yields
\begin{equation}
\langle  \{0\}_b, \tilde{N}_+ -1, \tilde{N}_- | c_{l,+} | \{0\}_b,
\tilde{N}_+, \tilde{N}_- \rangle = (-1)^{\tilde{N}_-} \delta_{l,\tilde{N}_+}.
\end{equation}
The factor $(-1)^{\tilde{N}_-}$ occurs because we have to
commute $c_{l,+}$ through the product of $N_-$ fermionic operators of
the left movers if we assume $m_0$ to be odd. We note that {\em
  no} such factor occurs for the corresponding matrix element of the
left movers. The calculation of the ground state to ground state
matrix element of $\tilde \psi_+(v)$ using (B.8) is simple as both
exponentials involving the boson operators can be replaced by the unit
operator and the matrix element is just
$c_{\tilde{N}_+,\tilde{N}_-}^+$. The comparison therefore yields
\begin{equation}
c_{\tilde{N}_+,\tilde{N}_-}^+(v)=(-1)^{\tilde{N}_-}e^{iv\tilde{N}_+}
\end{equation}
and $c_{\tilde{N}_+,\tilde{N}_-}^-(v) = e^{-i v \tilde{N}_-}$. Together
with (B.12) and the definition $\tilde{\cal N}_
     \alpha \equiv {{\cal N}}_\alpha - (m_0 + 1) {\hat 1}$
this implies
\begin{equation}
\hat{O}_+(v)e^{-i \tilde{{\cal N}}_+ v}
(-1)^{\tilde{\cal N}_-}
 |\{0\}_b, \tilde{N}_+,\tilde{N}_- \rangle = 
|\{0\}_b, \tilde{N}_+-1, \tilde{N}_- \rangle~.
\end{equation}
If we apply the operator $\hat{O}_+ (v)e^{-i
  \tilde{\cal N}_+
  v}(-1)^{\tilde{\cal N}_-} $ 
to the states in Eq. (B.10) and use
  again that $\hat{O}_+(v)$ commutes with the boson operators we obtain
\begin{equation}
\hat{O}_+(v) e^{-i \tilde{{\cal N}}_+ v} (-1)^{\tilde{{\cal N}}_-}
|\{m_l\}_b, \tilde{N}_+, \tilde{N}_- \rangle = |\{m_l\}_b, \tilde{N}_+
-1, \tilde{N}_- \rangle~.
\end{equation}
This shows that the operator $U_+ \equiv \hat{O}_+(v) e^{-i \tilde{{\cal
      N}}_+ v} (-1)^{\tilde{{\cal N}}_-}$ is {\em independent}
      of $v$ and given by
\begin{equation}
U_+ =~\sum_{\tilde{N}_+, \tilde{N}_-}~\sum_{\{m_l\}} |\{m_l\}_b,
\tilde{N}_+ -1, \tilde{N}_- \rangle \langle \{m_l\}_b,\tilde{N}_+,
\tilde{N}_- |~.
\end{equation}
It follows immediately that $U_+$ is {\em unitary}, i.e. $U_+
U_+^\dagger  = U_+^\dagger U_+ = \hat{1}$. From (B.15) one can infer 
that for arbitrary
functions $f$ of the number operator ${\cal N}_+$ one has
$U_+f(\tilde{{\cal N}}_+) = f(\tilde{{\cal N}}_+ + 1) U_+$. Concerning
the $\tilde{N}_+$-dependence the operator $U_+$ resembles a
one-dimensional tight-binding nearest neighbour hopping operator
\cite{AM}. Therefore the eigenstates of $U_+$ are just the corresponding
``Blochstates''
\begin{equation}
|\{m_l\}_b, k, \tilde{N}_- \rangle \equiv ~\frac {1}{\sqrt {2\pi}}
\sum^\infty_{\tilde{N}_+ = -\infty}~e^{i
k \tilde{N}_+}| \{m_l\}_b, \tilde{N}_+, \tilde{N}_- \rangle
\end{equation}
and the eigenvalue of $U_+$ is given by $e^{i k}$. If one defines the
operator
\begin{equation}
\hat{k}_+ \equiv~\sum_{\{m_l\}}~\sum_{\tilde{N}_-}~\int^\pi_{-\pi}~
|\{m_l\}_b,k, \tilde{N}_- \rangle k \langle \{m_l\}_b, k, \tilde{N}_- | d k
\end{equation}
the operator $U_+$ can be written as
\begin{equation}
U_+ = e^{i \hat{k}_+}~.
\end{equation}
In the ``$k$-representation'' the particle number operator $\tilde{{\cal
    N}}_+ $ acts like the differential operator $i \partial/\partial
    k$, which seems to imply the commutation relation $[
    \tilde{{\cal N}}_+,\hat k_+] =  i {\hat 1}$~.
This is {\em not} correct as an operator identity in the full
Hilbert space. In the $k$-representation the operator $ \hat k_+ $
acts as a multiplication operator which destroys the $2\pi$-periodicity
in $k$ unless the state $|\psi \rangle $ fulfills
$\langle  \{m_l\}_b,k=\pi ,\tilde{N}_- |\psi \rangle =0=
\langle  \{m_l\}_b,k=-\pi ,\tilde{N}_- |\psi \rangle $.

\noindent The direct calculation of the commutator using

\begin{equation}
     \tilde{{\cal
    N}}_+          =~\sum_{\tilde{N}_+, \tilde{N}_-}~\sum_{\{m_l\}}
 |\{m_l\}_b,
\tilde{N}_+ , \tilde{N}_- \rangle \tilde{N}_+ \langle \{m_l\}_b,\tilde{N}_+,
\tilde{N}_- |~.
\end{equation}
yields
\begin{equation}
[    \tilde{{\cal N}}_+,\hat k_+] =  i \left ( {\hat 1}
-2\pi \sum_{\{m_l\}}~\sum_{\tilde{N}_-}
|\{m_l\}_b,k=\pi, \tilde{N}_- \rangle  \langle \{m_l\}_b, k=\pi,
\tilde{N}_- |
 \right )
\end{equation}
in contrast to statements in the literature \cite{H}.

\noindent To summarize we have shown that

\begin{equation}
\hat{O}_+ (v) = U_+e^{i\tilde{{\cal N}}_+ v}
  (-1)^{\tilde{{\cal N}}_-} 
 =    e^{i \hat{k}_+} e^{i\tilde{{\cal N}}_+ v}
  (-1)^{\tilde{{\cal N}}_-} 
 \end{equation}
In the corresponding expression $\hat{O}_- (u)=  e^{i \hat{k}_-}
 e^{-i\tilde{{\cal N}}_- u}  $ {\it no}
 factor
$(-1)^{\tilde{{\cal N}}_+}$ appears and therefore $\hat{O}_+ (v)$ and
$\hat{O}_-(u)$ anticommute, which is necessary to enforce
anticommutation relations between $\tilde{\psi}_+(v)$ and
$\tilde{\psi}_- (u)$.
The bosonization of $c_{l,+}$ itself is obtained by the inversion
formula to (B.6)
\begin{equation}
c_{l,+} =~\frac{1}{2 \pi}~\int^{\pi}_{-\pi}~e^{-i v l} \tilde{\psi}_+ (v)
d v~.
\end{equation}
We finally show how the bosonization of a {\em single} fermion
operator described in this appendix relates to the bosonization of the
{\em product} $\tilde{\psi}_+(u)^\dagger \tilde{\psi}_+(v)$ described
in I. Using (B.8) and (B.21) we can write the product as
\begin{eqnarray}
\tilde{\psi}_+^\dagger(u) \tilde{\psi}_+(v) & = & e^{-i 
\phi_+^\dagger(u)}e^{-i
  \phi_+(u)} e^{i\phi_+^\dagger(v)} e^{i \phi_+(v)} e^{-i\tilde {{\cal
      N}}_+
(u-v)}
  \nonumber \\
& = & e^{-i (\phi_+^\dagger(u) - \phi_+^\dagger(v))} e^{-i(\phi_+(u) -
 \phi_+(v))}
  e^{[\phi_+(u), \phi_+^\dagger(v)]} e^{-i 
\tilde{{\cal N}}_+(u-v)}~.
\end{eqnarray}
Here we have used the Baker-Hausdorff formula $e^A e^B = e^B e^A
e^{[A,B]}$, valid if [A,B] commutes with A and B. The commutator in
the exponent is given by
\begin{eqnarray}
\left[ \phi_+(u), \phi_+^\dagger (v) \right] & = &
~\sum^\infty_{n,n'=1}~\frac{e^{inu}}{\sqrt{n}}~\frac{e^{-i n'
    v}}{\sqrt{n'}}~ [b_n,b_{n'}^\dagger ] =~\lim_{\eta \rightarrow 0}
\sum^\infty_{n=1} ~e^{in
    (u-v+i \eta)}/n \nonumber \\
& = & - \lim_{\eta \rightarrow 0} \left[\ln \left( 1 - e^{i(u-v
       + i \eta)}  \right)  \right]~.
\end{eqnarray}
In order to make the infinite series convergent we have added a
convergence factor $e^{-\eta n}$ with $\eta$ going to zero in the end
of the calculation. This is necessary as we had to perform the limit
$m_0 \rightarrow \infty$ in Eq. (B.6), i.e. we work with an
{\em infinite} Dirac see as in Luttinger's paper \cite{L}. Combining
 Eqs. (B.23) and (B.24) gives 
\begin{equation}
\tilde{\psi}_+^\dagger(u) \tilde{\psi}_+(v) = ~\frac{e^{-i\tilde{{\cal
        N}}_+(u-v)}}{1-e^{i(u-v+i 0)}}~e^{-i(\phi_+^\dagger(u) -
 \phi_+^\dagger (v))}~ e^{-i(\phi_+(u) -
 \phi_+(v))},
\end{equation}
which corresponds to Eqs. (47) and (49) of I.

A convergence factor also has to be introduced 
explicitely in the bosonization formula for the field operator (B.8)
if {\it no} normal ordering is used.
If one uses the Baker-Hausdorff formula
and (incorrectly) assumes the commutation relation
 $[ \tilde{{\cal N}}_+,\hat k_+] =  i {\hat 1}$~
one obtains
\begin{equation}
\tilde{\psi}_{\alpha}(u)=\lim_{ \eta \rightarrow 0}
\frac{1}{\sqrt { \eta}}e^{i \hat \theta_{\alpha}^{[\eta]}(u)},
\end{equation}
where the ``phase field'' operator
\begin{equation}
 \hat \theta_{\alpha}^{[\eta]}(u)\equiv \hat k_{\alpha}
+\alpha (  \tilde {{\cal
      N}}_{\alpha}  -\frac {1}{2})u +\phi_{\alpha}^{[\eta]}(u)+ 
(\phi_{\alpha}^{[\eta]}(u))^{\dagger}+\frac{\pi}{2}(1+\alpha) \tilde {{\cal
      N}}_{-\alpha},
\end{equation}
with
\begin{equation}
\phi_{\alpha}^{[\eta]}(u)
=-i\sum_{n \neq 0}\Theta (\alpha n)\frac{e^{inu}}{\sqrt{|
    n|}}~b_n~e^{-|n| \eta/2}
\end {equation}
contains the boson as well as the particle number changing
contribution. It is used in various applications \cite {V2} e.g the
discussion of transport.
It is obviously more rigorous 
to use of the normal ordered form (B.8), as always done in the present paper.


\begin{thebibliography}{99}
\bibitem{warning}
We should express a warning: Many authors in the present field,
  e.g. \cite{H}, use a
different sign convention in the exponent ($e^{ik_n x}\to   e^{-ik_n
 x}  $). This is in contradiction to the usual sign convention for a
plane wave in quantum mechanics.
\end{thebibliography}
\end{document}